# Tunable Magnetic Semiconductor Behavior Driven by Half-Filled One Dimensional Band in Zigzag Phosphorene Nanoribbons

Yongping Du[1,+], Huimei Liu[1,+], Bo Xu[2,*], Li Sheng[1], Jiang Yin[2], Chun-Gang Duan[3], Xiangang Wan[1,*]

[1]National Laboratory of Solid State Microstructures, and Department of Physics, Nanjing University, Nanjing, 210093, China

[2]National Laboratory of Solid State Microstructures and Department of Materials Science and Engineering, Nanjing University, Nanjing, 210093, China

[3]Key Laboratory of Polar Materials and Devices, Ministry of Education, East China Normal University, Shanghai 200062, China

**ABSTRACT**: An antiferromagnetic insulating state has been found in the zigzag phosphorene nanoribbons (ZPNRs) from a comprehensive density functional theory calculations. Comparing with other one-dimensional systems, the magnetism in ZPNRs display several surprising characteristics: (i) the magnetic moments are antiparallel arranged at each zigzag edge; (ii) the magnetism is quite stable in energy (about 29 meV/magnetic-ion) and the band gap is big (about 0.7 eV); (iii) a moderate compressive strain will induce a magnetic to nonmagnetic as well as semiconductor to metal transition. All of these phenomena arise naturally due to one unique mechanism, namely the electronic instability induced by the half-filled one dimensional bands which cross the Fermi level at around $\pi/2a$. The unusual electronic and magnetic properties in ZPNRs endow them great potential for the applications in nanoelectronic devices.

*Corresponding Author: xgwan@nju.edu.cn; xubonju@gmail.com.

[+]These authors contributed equally to this work.

Since the experimental realization of graphene nanoribbons by cutting graphene into nanometer-sized width[1], the electronic and magnetic properties of nanoribbon have raised a lot of attention[1-10]. Especially, the zigzag graphene nanoribbons (ZGNRs) were predicted to be antiferromagnetic semiconductors by noticing the two fold degenerated flat energy band at the Fermi level[2,3]. Moreover, it had been proposed that the half-metallicity in ZNGRs can be realized under an external transverse electric field[4], while many other methods, such as doping[5], defects[6] and edge-modification[7,8] were applied to tune or control the magnetism in ZGNRs. Partial theoretical predictions had been confirmed by the experiment[9,10]. In addition to graphene nanoribbon, magnetism had also been proposed for several other nanoribbons, such as BN[11], $MoS_2$[12] and ZnO[13]. However, for ZGNRs the electronic and magnetic properties strongly depend on the ribbon width[2], while the metallicity in magnetic $MoS_2$[12] and ZnO[13] nanoribbons limits their applications in nanoscale electronic devices.

Recently, a new two dimensional (2D) material, layered black phosphorus (phosphorene),[14-16] has been isolated in the laboratory through mechanical exfoliation from bulk black phosphorus. It was reported that phosphorene has carrier mobility up to 1000 $cm^2/V \cdot s$[14] and an appreciably high on/off ratio of 10000 in phosphorene transistor at room temperature[15], which makes phosphorene a potential candidate for future nanoelectronic applications[14-25]. Numerical calculation found that phosphorene is an insulator with direct bandgap, a 3% in-plane strain can change phosphorene to an indirect-gap semiconductor,[15] while a vertical compression can induce a semiconductor to metal transition.[17] The properties of phosphorene nanoribbons (PNRs), especially zigzag phosphorene nanoribbons (ZPNRs) and armchair phosphorene nanoribbons (APNRs) have also been studied.[20-23] It had been found that similar with phosphorene, the tensile stain and electric-field can modulate the physical properties of PNRs[20,22]. The possible structural reconstruction in the edge of PNRs had also been investigated[23].

Carefully inspecting the electronic band structure, we find that except those with very narrow width, ZPNRs always have two well-defined edge states. These edge states are exactly half-filled and cross the Fermi level almost at $k = \pi/2a$ ($a$ is the lattice parameter along the periodic direction of ribbon), consequently induce a giant instability as will be discussed later. We demonstrate that driven by this instability, the ground state of ZPNRs is antiferromagnetic (AFM) semiconductor. Compared with other magnetic nanoribbons, ZPNRs are semiconductors, their band gap, magnetic moment and the energy difference between the magnetic state and the nonmagnetic (NM) state are almost independent on the ribbon width. We also find that a moderate compressive strain will make the edge state crossing the Fermi level twice, which reduces the amplitude of susceptibility anomaly, thus results in a magnetic to nonmagnetic and semiconductor to metal transition. The stable and tunable electronic and magnetic properties in ZPNRs endow them great potential for the applications in nanoscale electronic devices.

We perform density functional calculations to study ZPNRs. For computational details, please refer to the Supporting Information (SI). The relaxed lattice constants in our HSE06 calculation for monolayer phosphorene are a = 3.29 Å, b =4.51 Å, in good agreement with other theoretical calculations.[15, 20] Following the convention about the GNRs[2,3], the ZPNRs are classified by the number of sawlike lines across the ribbon width as shown in Figure 1(a). It had been found that the edge states of the narrow ZPNRs are strongly hybridized together, we thus focus on the ribbons with width larger than 8, where the edge states are well defined, and the electronic structure and magnetic properties are almost not dependent on the width of ribbon. Upon structure relaxations, we find that the bond b1, which connects the edge P atom and P atoms in the interior of the nanoribbon as shown in Figure 1(a), decreases from 2.22 Å to 2.14 Å for 8-ZPNR. The corresponding edge angles α and β increase from 96.3° to 100.9°, and from 102.1° to 108.8°, respectively. Similar structural change has also been found for other PZNRs with different width, in consistent with the previous calculations.[20]

We perform non-spin-polarized calculation to check the basic electronic feature of ZPNRs. Consistent with other theoretical results[20], our HSE06 calculation also find that there are two bands crossing the Fermi level shown as the red lines in Figure 1(b).Those bands are quite separated from bulk ones and basically are pure edge states. It is interesting to see that those edge bands are quite narrow. For 8-ZPNR, the bandwidth is about 1.60 eV as shown in Figure 1(b), and the bandwidth will decrease slightly with increasing ribbon width as shown in SI2. Consistent with previous theoretical work,[20] our HSE06 calculations also find that the two edge bands are nearly degenerate around the zone boundary, and have considerable split at Γ point. Increasing the width of ZPNRs will reduce the interaction between two edges thus lead to the decrease of the edge state split at Γ point as shown in SI2. The most peculiar character of the electronic structure of ZPNRs with the width larger than 8 is that the well defined edge states are one-dimensional, exactly half-filled due to only one dangling bond per edge atom in the ZPNRs. The edge states are nearly degenerate in more than half of the Brillouin zone, consequently the half-filled band cross the Fermi level almost exactly at *π/2a* as shown in Figure 1(b) and SI2.

Here we demonstrate that such behavior of the edge state may bring exotic physical property to ZPNRs. Let us first study the famous Linhard function (LF) , which describes the bare electronic susceptibility and plays a crucial role in determining the structural and electronic properties of the system. In this expression, *f(k)* denotes the Fermi-Dirac distribution function, while $\varepsilon_k$ is the electron energy of the momentum $\hbar k$. It is well known that for 1D state, due to the effect of denominator, the susceptibility exhibits a large anomaly for $q=2K_F$, where $K_F$ is the wave vector on the Fermi surface[26]. In addition to the effect of denominator, we find that for ZPNRs, the numerator in LF also play important role[27]. Due to the requirement from *f(k)*, the nonzero contribution to the numerator can only come from those *k* and *k+q* in and out of the Fermi sea, respectively, and the integration of the numerator alone actually only depends on $K_F$ and *q*. For a given $K_F$, the maximum value of the

integration of the numerator in the LF (i.e., the available phase-space area) can be expressed as:

$$\max(\sum_k (f(k) - f(k+q))) = \begin{cases} 2K_F & (K_F < \frac{\pi}{2a}) \\ \frac{\pi}{a} & (K_F = \frac{\pi}{2a}) \\ \frac{2\pi}{a} - 2K_F & (K_F > \frac{\pi}{2a}) \end{cases}$$

Here we note that a too large value $k+q$ will again falls into the Fermi sphere due to the periodicity. When $K_F \sim \pi/2a$, as ZPNRs do, the integration of the numerator also reaches a maximum value at the same $q=2K_F \sim \pi/a$ point, where $\chi_q$ exhibits a singular behavior due to the effect of denominator. These two jointed effects result in abnormal behavior of $\chi_q$ at $q= \pi/2a$, which strongly suggests that there exists giant instability in ZPNRs at NM state.

To check the possible magnetic instability, we double the unit cell along the periodic direction, and perform the spin-polarized calculation for the ZPNRs with widths from 8 to 12. We select four magnetic configurations: ferromagnetic (FM), intra-edge FM and inter-edge AFM (AFM-1), intra-edge AFM and inter-edge FM (AFM-2), intra-edge AFM and inter-edge AFM order (AFM-3), as shown in SI3. We first perform calculation for 8-ZPNR, and find that FM and AFM-1 configurations are unstable and always converge to NM states finally. On the other hand, the intra-edge AFM configurations are energetically more favorable than the NM. We find that the AFM-2 configuration is the ground state, and is about 108.89 meV lower in energy than the NM state as listed in Table I. As shown in Figure 2(a), the magnetic moments are mainly located at the edge sites. The AFM-3 configuration is just 0.37 meV higher than the AFM-2 state, which indicates that the magnetic interaction between the two edges is very small and the large energy gain (about 109 meV) basically comes from the formation of AFM ordering along the edge.

We also calculate a series of N-ZPNRs (N from 9 to 12), and find that regardless the ribbon width, the FM and AFM-1 configurations are not stable. The energy difference between the AFM-2 and AFM-3 approach to zero with increasing the ribbon width, which again indicates that the inter-edge magnetic interaction is very weak. The AFM-2 and AFM-3 are almost degenerate in energy, and their electronic bands are also almost the same, thus the energy difference between the magnetic state and NM is an important value, which is close to 115 meV as shown in Figure 3(c). As shown in Figure 2(a), there are basically four magnetic sites, thus the energy gain per site is about 29 meV, which is higher than many other 1D cases.[11,12] As displayed in Figure 3, the magnetic moment at the edge atoms (around 0.155 $\mu_B$) also hardly depend on the ribbon width.

In addition to the magnetism, the anomaly of susceptibility may also result in a charge-density-wave (CDW) like distortion. We thus again double the cell along the periodic direction and relax both the lattice constant and internal coordinates of ZPNRs to check the possible structural reconstruction. We find that the ZPNRs maintain the original unit cell in the axis of ribbon and do not exhibit the Peierls like distortion. The possible edge reconstruction had also been discussed by Maity *et al.*[23], the energy gain of the structural distortion is only 0.6 meV/atom, which is close to the limits of density functional calculation[23], and also much less than the energy gain of magnetism as discussed at above. We thus believe the CDW transition will not occur, and there is no edge reconstruction for ZPNRs. In addition to the magnetic and structural transition discussed above, superconductivity is another possibility to remove the instability, we therefore invoke experimental efforts to study this possible rare 1D superconductor.

It is interesting to see that accompanying with the formation of AFM order at the edge, the edge bands split and the compound becomes a direct gap semiconductor with a sizable band gap (about 0.7 eV) opened for 8-ZPNR as shown in Figure 2(b). Moreover, we have also computed the band structures of a series of N-ZPNRs (N from 9 to 12), they are all semiconductor, and their band gaps are weakly dependent on the nanoribbon width as presented in Figure 3(a).

As pointed out above, the insulating AFM state of ZPNR arises from the strong instability induced by the 1D half-filled bands which cross the Fermi level at around

$\pi/2a$, thus it is natural to expect that the electronic and magnetic properties will change dramatically if one can tune the edge states. As is well known, the edge state is sensitive to doping, defect, external field, etc. For simplicity, here we only focus on the effect of strain. To analyze the effect of the compressive strains along the periodic direction, we display the electron structure of the edge state. As shown in Figure 4(b)(c)(d), consistent with the previous theoretical results[20], we find that the compressive strains has small effect on the bandwidth. Consequently, both the magnetic moment and the energy difference between the magnetic ground state and the NM state are not sensitive to small strains as shown in the Figure 4(a). The main influence of the compressive strains on the band structure is to enlarge the edge states splitting at $\Gamma$ point, and after a critical compressive strain (about 5%), the edge states will cross the Fermi level twice as shown in Figure 4(c). This will reduce the amplitude of the susceptibility anomaly at $q = \pi/a$. Consistent with that, our calculations show that now the ground state change from AFM semiconductor to nonmagnetic metal. We also perform calculation for other ribbon width, and find that regardless the width, a moderate strain(around 5%) always induce a magnetic to nonmagnetic and semiconductor to metal transition. In addition to the instability around $q = \pi/a$, basically we also need to check the instability induced by the Fermi surface around $\Gamma$ point. Unfortunately, as shown in Figure 4(b), this point is far from any simple fraction. Such incommensurate-like phase currently is beyond our ability to handle. We, however, believe this kind of ordering is unlikely to occur, due to the fact that the amplitude of $\chi_q$ at this point is smaller than that at $q = \pi/a$.

Due to the above discussed unique mechanism, the magnetic properties in ZPNRs are quite different with other magnetic nanoribbon systems. The magnetism of ZPNRs arises from the sites located at two edges, and along the edge the magnetic moment are AFM ordered. Whereas the edge is FM ordered for ZGNRs[2], the magnetic moment in zigzag ZnO nanoribbon is only contributed by the oxygen edge[13], and for zigzag MoS2 ribbon the magnetic moments of both inter- and intra-edge are FM coupled[12]. In addition, ZPNRs have following advantages: (i) the energy difference between magnetic ground state and NM state is around 29 meV/magnetic-site, which is substantially larger than that of ZGNRs and MoS2 nanoribbon. The rather stable magnetism in ZPNRs is very important for the electronic devices application at room temperature; (ii) a moderate in-plane compression, possibly caused by epitaxial mismatch with a substrate, can induce a sharp magnetic to nonmagnetic as well as a semiconductor to metal transition.

In summary, we have presented a detailed study on the intrinsic magnetic properties of ZPNRs. We find that due to the unique mechanism of magnetism, the ZPNRs exhibit several interesting properties: The ground state of ZPNRs is AFM semiconductor. The magnetism is rather robust, and the magnetic moments locate at the edge sites and are AFM ordered along the edge. The electronic and magnetic properties are only weakly dependent on the ribbon width. A moderated strain can induce a magnetic to NM and semiconductor to metal transition. The stable and tunable magnetic properties of ZPNRs make them great interest for further experimental investigation.

**ACKNOWLEDGMENT**

We acknowledge the useful discussions with S.Y. Savrasov. We acknowledges support by National Key Project for Basic Research of China (Grant No. 2011CB922101, 2013CB922301, 2014CB921104), NSFC(Grant No. 11374137, 91122035, 11204123, 11174124 and61125403) and PAPD.

**Table 1 total energy and magnetic moment of different magnetic configurations of 8-ZPNR.**

| 8-ZPNR | NM | FM | AFM-1 | AFM-2 | AFM-3 |
|---|---|---|---|---|---|
| Energy(meV) | 108.89 | 108.89 | 108.89 | 0.0 | 0.37 |
| Magnetic moment(μB) | 0 | 0 | 0 | 0.151 | 0.152 |

# Figure Captions

Figure 1: (a) Ball-and-stick model of 8-ZPNR; (b) Band structure of 8-ZPNR calculated by HSE06 scheme.

Figure 2: (a) Spatial spin distribution of 8-ZPNR, the red one denote the spin up and green one is the spin down with the isosurface value 0.002 e/Å$^3$; (b) Band structure of 8-ZPNR with AFM-2 configuration.

Figure 3: Electronic and magnetic properties of a series of N-ZPNRs(N=8 to 12): (a) band gap and (b) magnetic moment of ZPNR with AFM-2 configuration as the function of ribbon width; (c) energy differences between the NM state and AFM-2 ground state change with the ribbon width.

Figure 4: Effect of compressive strain on the magnetic and electronic properties of ZPNRs. (a) The dash line with black triangle is the energy difference between NM state and AFM-2 state, and the magnetic moment is shown by red line with solid circle; (b) (c) (d) are the band structures of 8-ZPNR under the compressive strain of 10%, 5%, 0%, respectively.

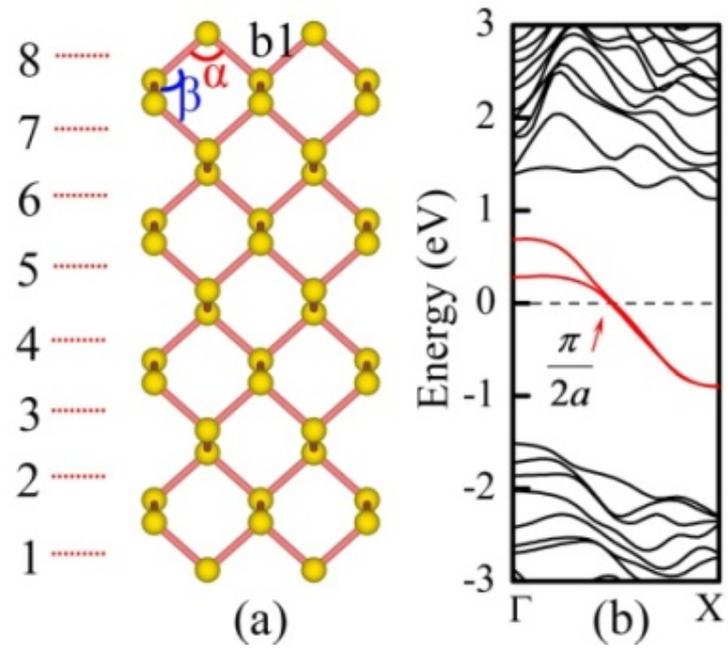

Figure 1

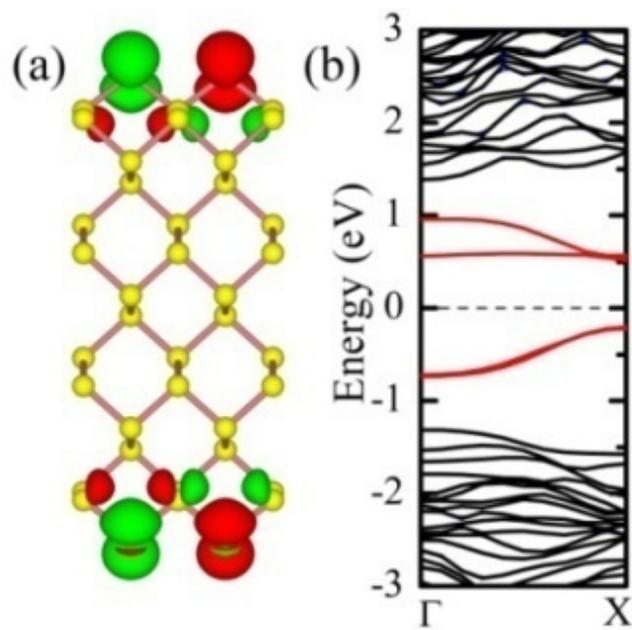

Figure 2

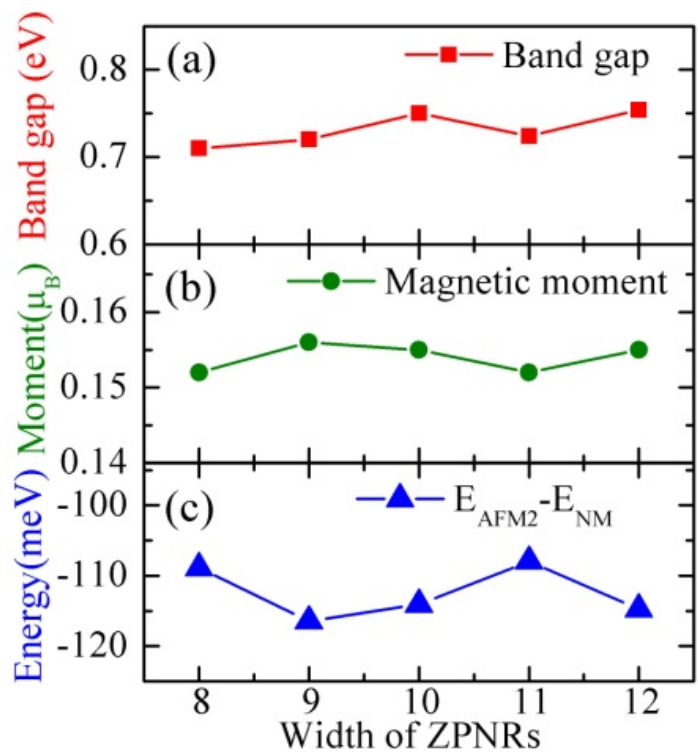

**Figure 3**

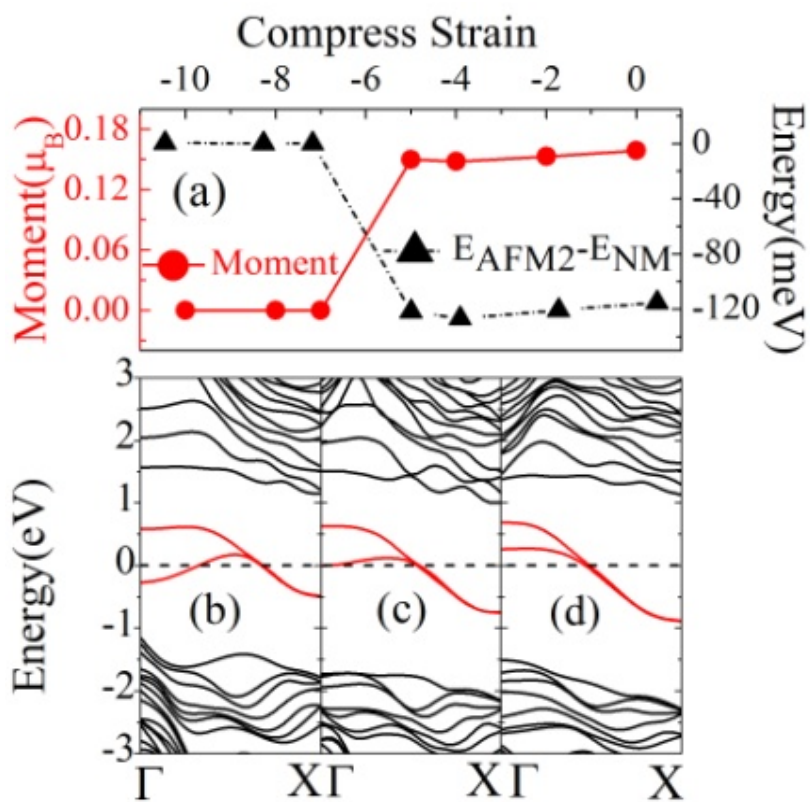

**Figure 4**

# Supporting Information 1 (SI1)

## Computational detail

Our density functional calculations were based on the Vienna ab initio simulation package (VASP)[1,2]. An energy cutoff of 500 eV was adopted for the plane-wave expansion of the electronic wave function and the energy convergence criteria was set to 10-5 eV. To sample the Brillouin zone, appropriate k-point meshes of (9*1*1) were used for calculations. Structures were relaxed until the force on each atom was less than 0.01 eV/ Å. It is well known that the local (local density approximation) or semi-local (generalized gradient approximation) approximations for the electronic exchange and correlation fail to cancel the self interaction error (SIE)[3]. On the other hand, with the correction of the SIE, hybrid function scheme had been shown are superior to the LDA and GGA in description of not only the lattice structure but also the electronic and magnetic properties[4-8]. Thus in this work, we apply the hybrid function of HSE06 [7,8] to investigate the electronic band structure and magnetic properties of phosphorene nanoribbon. A vacuum spacing of 18 Å is used so that the interaction in the non-periodic directions can be neglected.

## Supporting Information 2 (SI2)

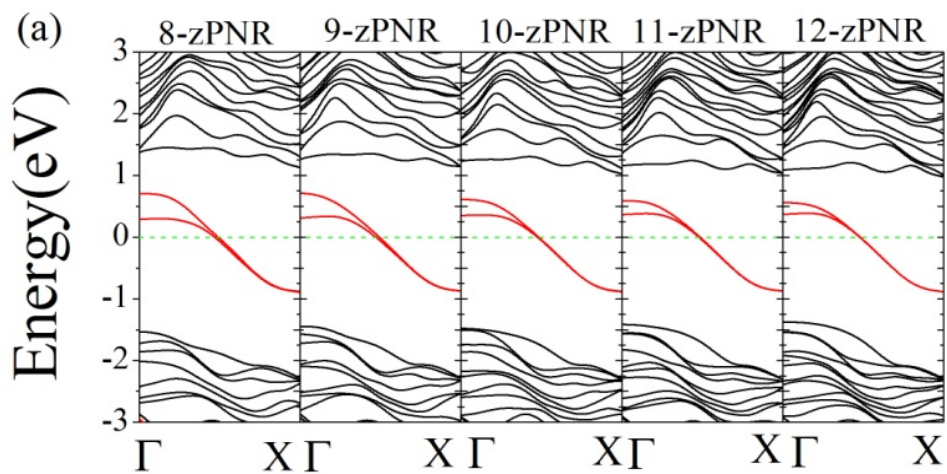

SI2 (a) The band structure of ZPNRs, calculated by HSE06 scheme.

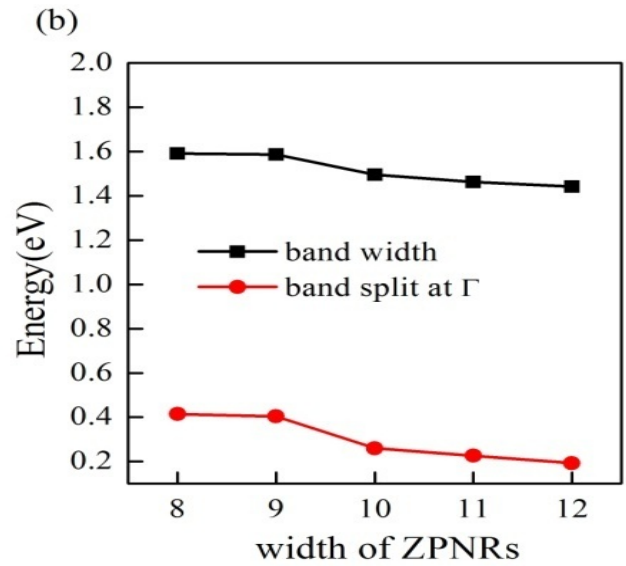

SI2 (b) Back line denotes the band width as the function of ribbon width, red line is the band split at the Γ point as the function of ribbon width.

## Supporting Information 3 (SI3)

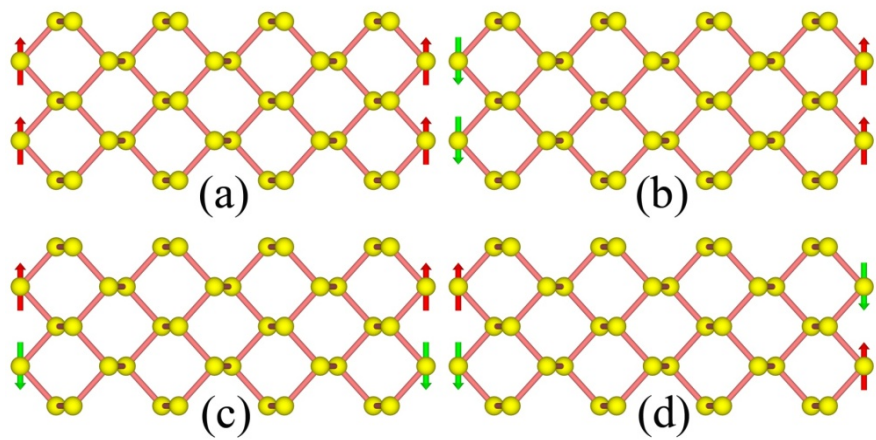

SI3 The initial magnetic structures adopt for searching ground state (a) FM order; (b) AFM-1 order; (c) AFM-2 order; (d) AFM-3 order.